**How to cite**

# Bringing active learning, experimentation, and student-created videos in engineering: A study about teaching electronics and physical computing integrating online and mobile learning

**Jonathan Álvarez Ariza**
Department of Technology in Electronics, Engineering Faculty, Corporación Universitaria Minuto de Dios-UNIMINUTO, 111021, Bogotá, Colombia
**Email: jalvarez@uniminuto.edu**

**Abstract:**

Active Learning (AL) is a well-known teaching method in engineering because it allows to foster learning and critical thinking of the students by employing debate, hands-on activities, and experimentation. However, most educational results of this instructional method have been achieved in face-to-face educational settings and less has been said about how to promote AL and experimentation for online engineering education. Then, the main aim of this study was to create an AL methodology to learn electronics, physical computing (PhyC), programming, and basic robotics in engineering through hands-on activities and active experimentation in online environments. N=56 students of two engineering programs (Technology in Electronics and Industrial Engineering) participated in the methodology that was conceived using the guidelines of the Integrated Course Design Model (ICDM) and in some courses combining mobile and online learning with an Android app. The methodology gathered three main components: (1) In-home laboratories performed through low-cost hardware devices, (2) Student-created videos and blogs to evidence the development of skills, and (3) Teacher support and feedback. Data in the courses were collected through surveys, evaluation rubrics, semi-structured interviews, and students' grades and were analyzed through a mixed approach. The outcomes indicate a good perception of the PhyC and programming activities by the students and suggest that these influence motivation, self-efficacy, reduction of anxiety, and improvement of academic performance in the courses. Even, the PhyC activities can reduce the differences in academic performance between women and men students and promote their improvement towards excellence with better results for courses combining online and mobile learning. Similarly, women students prefer to use more often tutoring services and had a better perception of the teacher's feedback than their male pairs. The methodology and previous results can be useful for researchers and practitioners interested in developing AL methodologies or strategies in engineering with online, mobile, or blended learning modalities.

1. Introduction

Active Learning (AL) is a teaching method with special educational features such as being student-centered, active participation of the student in developing hands-on tasks, planning the students' work according to defined learning outcomes, peer collaboration, and the role of the professor as a mentor and facilitator of the learning process [1]. These features are aligned with the learning outcomes and the skills that are developed and expected in the students in the engineering curricula, particularly in this study, in electronics, programming, and Physical Computing (PhyC). Typically, in engineering education, AL has been incorporated into methodologies such as problem-based learning, project-based learning, cooperative-based learning, or team-based learning, among others [1]. Several prior studies illustrate the advantages and challenges to include AL in engineering curricula from undergraduate to master's degrees. For instance, Vodovozov et al. [2] show a study for the courses of robotics and industrial automation and drives. The authors expose several alternatives to engage the students in AL as optional exercises and labs, presentations, lectures, and demonstrations. Nonetheless, this engagement depends on the learning styles, the time destined for the activities by the students, and the feedback of the instructors. Indeed, other conditions such as working in small groups, individual instruction methods, and problem-based or project-based learning must be given in the context in which AL is implemented. In [3] is presented a novel framework for AL according to the mapped educational outcomes. In the study, the teacher or instructor should prepare a course based on Bloom's taxonomy, measuring the course outcomes attainment using feedback, and mapping the outcomes expected directly to the taxonomy levels. Miranda et al. [4] point out the features of AL for the concept of "education 4.0" in electrical engineering, which includes the understanding of user/customer requirements, acquiring and analyzing data, or creating prototypes. Education 4.0 requires active and higher independence of the learner, ICT tools platforms powered by the Internet of Things (IoT), and the role of the educator as mentor, collaborator, or facilitator of the learning process. In addition, the authors comment that AL methodologies have been achieved mainly in face-to-face learning environments. Similarly in [5], the researchers depict a problem-based learning methodology based on steps such as the development of the experience through planning and preparation, the initial explanation of the theoretical concepts, the allocation of groups, projects, and roles, and continuous monitoring and evaluation. Among the skills most developed by students are self-learning, knowledge integration, critical capacity, and teamwork. The authors remark on the importance of the methodology to develop the skills required in the business world such as critical capacity, communication ability, planning ability, autonomy, and teamwork. Also, a summary of practices in engineering education for modeling and simulation that includes AL is illustrated in the study [6]. According to the authors, the active learning approaches mentioned in the study can provide instructors with guidance on how to implement pedagogies that foster meaningful learning. However, the authors state that is needed more studies that report the integration of the previous practices for engineering education in the fields of modeling and simulation. At last, Desai [7] describes the advantages to incorporate AL in engineering education using techniques such as Jigsaw in which a topic is provided to the students that discuss and share their learning that is monitored by an expert group or the case studies in which students can visit an industry or workshop to identify industrial and social applications.

Regarding PhyC, it has been incorporated into the engineering curricula since the last decade to benefit the development of skills of the students in computational thinking (CT), tinkering, robotics, algorithmic thinking, peer-to-peer collaboration, and promote self-efficacy and motivation in diverse academic programs from elementary schools to higher education institutions (HEIs) [8–11]. From a conceptual perspective, physical computing is a term coined for a combination of hardware and software to build physical and tangible systems that sense and interact with the real world [12,13]. PhyC needs interaction with algorithms and artifacts such as sensors, 3-D printing devices, actuators like servomotors for robotics, or displays to show sensors' information, among others. Other authors such as O'Sullivan and Igoe [14] indicate that PhyC is an interaction, where the term *interaction* is understood as "an iterative process of listening, thinking, and, speaking between two or more actors". PhyC involves the co-creation of physical artifacts with programming and making activities that promote learning and engagement of the students because they can experiment and apply their learning and knowledge in the creation of devices such as robots and can share their experiences with others, which fosters collaboration and interaction between peers [15]. All these advantages of PhyC require AL, experimentation, and hands-on activities by part of the students, but also require the presence of the teacher as a guide and scaffolder for their educational process. From there, the educational tenets of PhyC are based mainly on constructionism, a learning theory developed by Seymour Papert [16–18]. Its foundations are based on constructivism and experimentation through hands-on activities made by the students in the field of programming and computer interaction [19]. Constructionism deals with the idea that knowledge and learning should be created by the students instead of being a simple act of information transmission between teachers and learners. Students learn better when they can construct tangible things, and they can see the effect of their algorithms or programs to create generalizations and abstractions [15,20].

While the previous aspects and studies illustrate the advantages to include AL and PhyC in the curricula of engineering for learning and engagement of the students, they have been achieved principally in face-to-face learning environments and less has been said on how to develop AL and PhyC in online or blended learning modalities. This is a particularly difficult issue because one of the requirements of the engineering curricula and PhyC is active experimentation and hands-on activities which in entirely online learning settings is a limitation due in part to its features and the type of computer-mediated communication thereof. Clearly, this issue was increased by the context generated by the COVID-19 pandemic around the world which accelerated the digitalization process of education with the migration of programmatic contents of the courses to online platforms like Moodle or similar [21,22]. Then, teachers and students who had been working in a face-to-face learning modality were quickly turned to an online learning modality without many tools at least from a pedagogical standpoint [23,24]. As Zawacki-Richter points out [24], the process of digitalization of education was more assimilated and adequate by humanities, cultural studies, linguistics, and natural sciences, but for areas that have a huge portion of laboratory work, for instance, engineering or technology academic programs, the transition to digital learning was *hardly feasible* due to the complexity, time-consuming, and expensive of this process for HEIs. However, it is worth mentioning that the process of digitalization and transformation of face-to-face courses to online or blended learning modalities started before the pandemic in several HEIs around the world in the field of engineering or computer science [25–28] and it was reenforced during the pandemic [29–31]. Another factor that is important to mention is that during the pandemic the face-to-face courses were quickly adapted to a format of emergency remote teaching (ERT) [21]. ERT differs from online learning since it is a temporal method of learning and teaching due to crisis circumstances. As Hodges et al. [32] describe, the primary objective of the ERT is not to recreate a robust educational system but to be a method to support the learning needs of the students during the crisis period.

In this sense, the main aim and motivation of this study were to create an active learning methodology to learn electronics, physical computing, basic robotics, and programming in engineering through hands-on activities and active experimentation in online environments and not merely an ERT methodology. Students in online engineering courses in electronics or electrical fields that combine hands-on activities could struggle with the form to apply the concepts learned and this can be a cause of dropout in their careers. Even, during the COVID-19 pandemic, we received a lot of students' comments regarding the way to effectively apply the concepts with laboratories and practical activities. Also, the initiative arose by the lack of studies in the field of engineering and online learning that promote effectively active learning and experimentation and not only the delivery of content and knowledge to the students during the COVID-19 pandemic and even before it. The methodology was created employing the guidelines of the Integrated Course Design Model (ICDM) [33] that contemplates three components: defining learning goals, defining teaching, and learning activities, and feedback and assessment, which are aligned with the prior studies described above. These components provided a perspective to design a PhyC course with AL as an instructional method considering the context of the students, learning outcomes, learning activities, feedback, and the role of the instructor.

Thus, the methodology was carried out during the years 2020-2021 with 56 students in two engineering programs: Industrial Engineering and Technology in Electronics. For that, several hardware devices such as an Arduino board, multimeter, protoboard, resistances, and a robot in 3-D printing were sent to the students' home. In addition, in some courses was employed a mobile learning app to allow ubiquity in the learning process. Hence, in these courses, online and mobile learning were combined to provide a better learning experience for the students. These technological tools in hardware and software served to perform the different activities in the courses combining electronics, PhyC, and programming. Thus, the home spaces were transformed into AL environments to understand and apply the topics and concepts viewed in the courses, meanwhile, the online classes served to clarify the doubts and provide feedback to the students. It is worth mentioning that the results presented in this study expand the methodology and findings of the work in [23]. In addition, the rationale to create the methodology was threefold. Firstly, there was an important limitation of the laboratories and hands-on activities that the students could perform due to the lockdowns and mobility restrictions during the years 2020-2021. Thus, the activities inside the courses with laboratory support were strongly impacted by the COVID-19 pandemic and were replaced mainly by the usage of simulators, remote laboratories, augmented reality, mobile learning, or virtual environments [34–39], but not for all HEIs due to their costs. Thus, this situation allowed the creation of a methodology to support the active experimentation of the students mixing in-home work with online classes employing low-cost hardware components that did not represent excessive cost overrun for the HEI in which the methodology was carried out.

Secondly, the COVID-19 pandemic was a big concern for many students around the world due to the socioeconomic panorama produced by it [40,41]. Several students in the context of this study manifested the intention to drop out of the educational process at the university. Then, the methodology searched to increase motivation and strengthen self-efficacy in the courses. In special, self-efficacy has been reported by some studies as an important factor that could affect the performance and attitudes in learning activities and dropout rates, and it is affected by physical and emotional factors [42–44].

Thirdly, one important debate has arisen during the COVID-19 pandemic and in the post-COVID-19 period on how to evaluate the learning outcomes of students in online or blended learning environments. Especially in engineering, only exams in online platforms with closed-ended questions rather than hands-on activities could not give enough information about the learning process followed by a student. This issue in online learning settings has been studied and manifested by several researchers [45,46]. Researchers in these references mention that a *common practice* is the efficient delivery of content in form of lectures, quizzes, or exams in online courses in learning management systems (LMSs) instead to offer the students a constructivist approach where collaboration, dialog, and development of critical thinking are fostered. Thus, apart from the mentioned technological tools, the methodology included the usage of web 2.0 tools (blogs and videos) [47] with the aim to the students collaborated and shared information with their peers, as well as follow up their educational process by the teacher in the comprehension and application of the concepts viewed in class. Each blog created by the students in groups was assessed by an evaluation rubric whose structure was reviewed by an external education expert. Then, the students were co-creators of digital media which has been a strategy to demonstrate learning and engagement in several knowledge areas [48–50]. These three aspects synthesize the rationale and main components of the online learning methodology with physical computing presented in this study. Under these components, the research questions (RQs) defined for the study were the following:

**RQ1**. What are the students' perceptions regarding learning, experimentation and motivation with the in-home labs and the creation of blogs and videos in the online courses?

**RQ2**. What factors influence active learning, experimentation, and motivation in the courses combining online and mobile learning?

**RQ3**. What are the educational implications of the methodology for the academic performance of the students?

According to the mentioned, the remainder of the paper is divided as follows. Section 2 describes the educational context in which the methodology was developed and describes the main components of the methodology altogether with the instruments to collect data and its analysis. Section 3 depicts the results based on the RQs proposed and contrast some of these findings with other prior studies. Section 4 discusses the implications of the methodology for practice in engineering education. Section 5 exposes the limitations of the study. Finally, section 6 outlines the conclusions and final remarks of this study.

## 2. Method and Materials

### 2.1 Context and participants

The methodology was carried out during 2020 and 2021 in two engineering programs of the Colombian HEI Corporación Universitaria Minuto de Dios-UNIMINUTO: Industrial engineering and Technology in Electronics. To deploy the methodology, it was chosen the subjects of Automation and Introduction to Electronics offered by these programs respectively. These subjects were selected because they offered the first approach of the students to electronics and programming topics. Besides, the students of Introduction to Electronics are in the first semester of the program of Technology in Electronics, whereas the students of Automation are in the eighth semester of Industrial Engineering. N=56 students decided to participate voluntarily in the methodology, which counted with the characteristics of gender and age depicted in Table 1. In addition, this table presents the schedule in which the students took the classes in synchronous online mode.

**Table 1**. Participants' characteristics. I, II indicate the semester of the year.

| Subject | Period | Gender radio | Schedule | Age average |
|---|---|---|---|---|
| Introduction to Electronics | 2020-I | 15 students: 100% male | Night (8:30pm-10:00pm) | 20.93 years |
| Automation | 2020-II | 26 students: 53.84% male, 46.15% female | Night (8:30pm-10:00pm) | 24.52 years |
| Automation | 2021-I | 15 students: 73.33% male, 26.66% female | Morning (9:15am-10:45am) | 23.86 years |

In several Colombian HEIs is common that the students can take their subjects in three schedules: morning, afternoon, and night. Then, the methodology was deployed in the courses both in the night and in the morning schedules. Typically, students that prefer a night schedule work during the day and take their subjects in the schedule from 6:00 pm to 10:00 pm. Also, each semester is divided into 16 weeks with three terms of 5 weeks approximately. Concerning the topics in the courses, the subject of Automation is focused on the fundamentals of electric circuits, industrial sensors, simulation, and programming of hardware devices such as Arduino, servomotors, or sensors. The subject of Introduction to Electronics is focused on electric circuits, fundamentals of analog and digital sensors, algorithms for embedded systems, simulation, and design of Printed Circuit Boards (PCBs).

**2.2 Educational methodology**

The methodology was outlined according to the criteria and guidelines of the *Integrated Course Design Model* (ICDM) provided by Fink [33] to create significant learning experiences. This model articulates three overall components: learning goals, assessment and feedback, and teaching and learning activities. The overall goal of this model is to create significant learning experiences through AL, and the integration and reflection on the previous components in the teaching-learning process. Each component inside the methodology is exposed as follows:

*Defining learning goals*: A first step to create the methodology was to design a criterion of the outcomes regarding the knowledge and skills developed by the students in the areas of electronics, PhyC, and programming. To meet this requirement, it was constructed a matrix with the topics in each term of the semester, the expected students' outcomes, and the taxonomy level according to Marzano's taxonomy [51]. The developed matrix is available in Table 2. Marzano's taxonomy was proposed to respond to the shortcomings of Bloom's taxonomy, in special, the interaction between the logical and empirical aspects that can be involved in the construction of learning [51]. Marzano's taxonomy complements Bloom's taxonomy concerning *knowledge utilization* including what is defined as mental processes (decision-making, problem-solving, experimenting, investigating), which is appropriate for hands-on activities in engineering. Given that the methodology gathers active learning through experimentation with hands-on activities, the students can perform different types of tasks, for instance, from the utilization of Ohm's law to solve an electrical circuit up to the creation and implementation of an algorithm or code to move a robot.

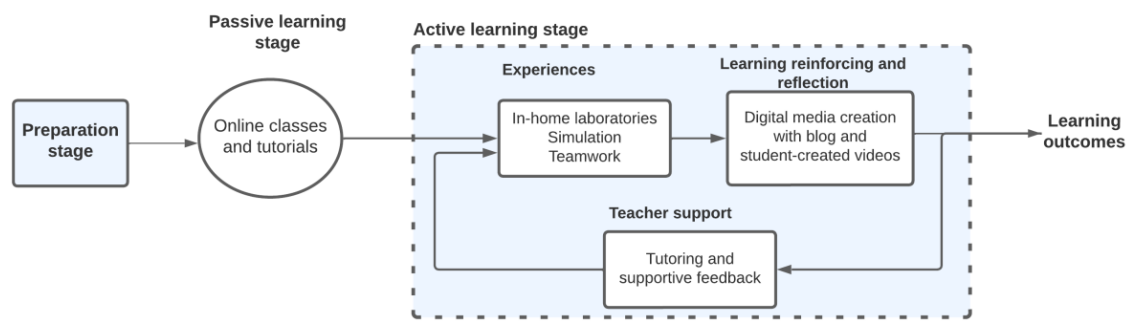

**Figure 1.** AL model tailored for the methodology according to the ICDM.

**Table 2**. Learning outcomes expected by the students in the courses by term.

| Term (FT-TT) | Topics | Students' outcomes (Based on Marzano's Taxonomy) | Marzano's Taxonomy Level |
|---|---|---|---|
| **1. FT: Fundamentals of electric circuits and instruments** | -Concept of voltage, current, and resistance.<br>-Ohm's law<br>-Series, parallel, and mixed circuits<br>-Circuit Simulation<br>-Handling of multimeter and power supply<br>-Handling of protoboard to construct basic circuits | 1.1 The student **uses** Ohms' Law to find voltage, current, or resistance in an electrical circuit and comprehends the implications of these values in it.<br>1.2 The student **describes** the concepts of voltage, current, resistance, and power, and exemplifies cases of systems where these are used.<br>1.3 The student **simulates** the basic circuits (series, parallel, and mixed) and contrasts his/her results with the theory.<br>1.4 The student **uses** power supplies and a multimeter to perform basic electrical circuits with their measurements. | **Level 1: Retrieval (Executing, recalling)**<br><br>**Level 2: Comprehension (integrating)**<br><br>**Level 3: Analysis (Matching)** |
| **2. ST: Introduction to physical computing and programming** | -Introduction to algorithms for embedded systems and physical computing<br>-Fundamentals of analog and digital sensors,<br>-Handling of displays to visualize sensors' information<br>-Flow diagrams<br>-Implementation and simulation | 2.1 The student **solves** a problem with physical computing creating an algorithm to meet the design requirements.<br>2.2 The student **experiments** to validate and test his/her physical computing systems.<br>2.3 The student **symbolizes** his/her algorithms with flow diagrams and **comprehends** their structure.<br>2.4 The student **identifies errors or problems** in the algorithms and hardware components and proposes a solution to fix them. | **Level 2: Comprehension (Symbolize)**<br><br>**Level 3: Analysis (Analyzing)**<br><br>**Level 4: Knowledge Utilization (problem-solving, experimenting)** |
| **3. TT: Integrating physical computing, programming, and basic robotics.** | -Algorithms for embedded systems, basic robotics, and physical computing<br>-Design of Printed Circuit Boards (Only for students of introduction to electronics)<br>-Soldering ( Only for students of introduction to electronics)<br>-Basics of Hardware integration and testing | 3.1 The student **investigates and selects** suitable hardware components to meet the design requirements of the project or problems proposed.<br>3.2 The student **develops** a solution for the project or problem according to the physical computing, and basic robotics requirements in hardware and software.<br>3.3 The student **makes experiments** that allow checking the functioning of the solution for the indicated project or problems. | **Level 4: Knowledge Utilization (problem-solving, experimenting, investigating, decision-making)** |

Then, Marzano's taxonomy provides a suitable set of activities and behaviors that are expected by the students when combining different levels of difficulty and knowledge application in their activities in the levels of comprehension, problem-solving, analysis, and metacognition, among others [51]. For each course, the learning outcomes were defined in function of the topics of each term. Thus, for both subjects (Introduction to Electronics and Automation), the first term (FT) entailed the fundamentals of electric circuits and instruments, the second term (ST) addressed introduction to physical computing and programming, and the third term (TT) encompassed integration of physical computing, programming, and basic robotics. Besides, these topics allowed to plan the in-home laboratories, the software used, and the hardware materials that the students would employ.

*Teaching and learning activities*: Once the learning outcomes were defined, the teaching-learning activities were posed to respond to the educational needs of the students in the online environment with the incorporation of AL. These activities were structured according to Figure 1 in function of the ICDM.

The teaching-learning activities were divided into two stages: passive and active. In the passive stage, the students received their classes in a synchronous mode according to the schedule of Table 1 using mainly Microsoft Teams. The classes were recorded and uploaded to the LMS Moodle for the students to access at any time to review the concepts and knowledge addressed. So, in this stage, the students received information and ideas in a passive form. With this information, the students passed to the active learning stage in which a set of problems and tasks were provided through assignments and laboratories. The problems gathered theoretical components but with an important load of hands-on activities with the in-home laboratories. To perform these laboratories, several low-cost hardware devices were sent to the home of each student, such as Arduino, a robot in 3-D printing, a multimeter, a protoboard, resistances, among others. With the indications in the classes and the tutorials created to handle the laboratories, the students experimented directly in the home spaces. Also, the students used simulators to contrast their results in the problems with the theoretical concepts and hands-on activities. Once the students constructed experiences, they created digital media in form of videos and a blog with the solution of assignments, exams, or laboratories. Especially, the videos were focused on *how* the students applied the knowledge acquired and connected this with the in-home laboratories. The blog and videos had the following criteria:

1. The students should work in groups of two or three people.
2. Each assignment, project, or exam must count with an explanation video. Each student of the group at least must make a video by activity that evidences the application of the learned concepts with the help of the materials sent to the in-home laboratories. Also, an explanation of simulations was mandatory. The duration of the videos should oscillate between 3-8 mins.
3. The videos and the development of the different tasks in the courses must be uploaded to a blog created on any platform such as WordPress, Blogger, or Google sites. The videos could be uploaded to YouTube.
4. The blog must contain organized entries for the activities developed according to the topics viewed in each term.

*Feedback and Assessment*: Each video made by the students and the blog was assessed under the lens of the learning outcomes defined in the courses. After, feedback was provided to the students according to the errors, problems, or misconceptions in the topics addressed. The feedback was provided to the student to recognize the problems, but also providing the opportunity to fix them, debug the found errors, and experiment directly with the in-home laboratories. Feedback was provided timely and efficiently using the classes, WhatsApp, emails, and the LMS system with Moodle. Additionally, traditional exams were replaced by problems that required experimentation, extending the time of delivery for those exams. With this form of assessment, the students integrated the concepts learned, the application thereof, and the reflection on the topics in the classes. This form of exam arose because an approach focused on the delivery of content with multiple choice exams or written exams cannot provide enough information about what the students are learning and applying in engineering. Besides, feedback and assessment are related to the pedagogical approach adopted by the teacher or instructor and have a relationship with the cheating behaviors that could be exhibited by the students in online exams [52].

**2.3 Materials**

2.3.1 *Hardware*

To perform the different in-home laboratories with electronics, PhyC, and programming, several low-cost hardware components were selected in the preparation stage of Figure 1 that allowed the students to make the different hands-on activities to foster the learning goals in the courses. The materials were sent to the home of the students as *laboratory kits* with the components illustrated in Table 3. An example of the kits is depicted in Figure 2. Once the semester finished, a face-to-face class session was planned for the return of all components. Each kit had an approximate cost of USD 40 and was delivered to the students directly by the university. A total of 30 kits were constructed to offer experimentation with the in-home laboratories because this is the maximum of students allowed by the university in a course.

**Table 3.** List of hardware components of each kit employed by the students in the in-home laboratories.

| Component | Quantity | Course topic |
| --- | --- | --- |
| Resistances 220 Ω, 330Ω, 510Ω, 1KΩ, 2KΩ, 10KΩ | 20 | Fundamentals of electric circuits, PhyC |
| Multimeter | 1 | Fundamentals of electric circuits, handling of laboratory instruments, PhyC. |
| Arduino UNO-Arduino compatible board | 1 | PhyC, programming |
| Protoboard | 1 | Fundamentals of electric circuits, PhyC, programming |
| Temperature sensor (LM35) | 1 | PhyC, programming |
| Liquid Crystal Display (LCD)-2x16 | 1 | PhyC, programming |
| Leds | 10 | PhyC, programming |
| Robot in 3-D printing with ultrasonic sensor | 1 | PhyC, fundamentals of robotics, programming |
| Buttons (switches for protoboard) | 4 | PhyC, programming |
| 12V DC Adapter | 1 | Fundamentals of electric circuits, handling of laboratory instruments, PhyC, fundamental of robotics |
| Protoboard jumpers (male-male, male-female connectors) | 20 | Fundamentals of electric circuits, handling of laboratory instruments, PhyC, fundamental of robotics |
| Bluetooth Module (HC-05) | 1 | PhyC, programming, fundamental of robotics |

Specifically in the courses of 2020-I and 2021-I according to Table. 1, it was used an app known as *EmDroid* for PhyC activities as described, while for the course of automation of 2020-II was selected only the Arduino platform. In the courses that employed the app, the Arduino board was replaced by a compatible one with a Bluetooth module incorporated because the app employs this protocol to program the microcontroller in the development board with the code created by the students through graphical blocks. The next section explains the details of this app.

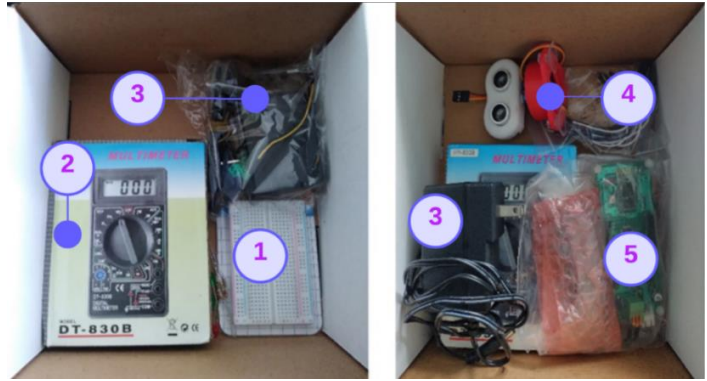

**Figure 2.** Example of the hardware materials sent to the students. 1. Protoboard, 2. Multimeter, 3. 12V DC adapter, 4. Robot in 3-D printing, 5. Compatible Arduino development board and LCD.

2.3.2 *Software*

To perform the different physical computing activities were selected Arduino IDE and an *app* called *EmDroid*. This last software is a proprietary app developed in 2017 in the OS Android by the program of Technology in Electronics at the UNIMINUTO university to create different ways in which students can learn PhyC by taking advantage of their smartphones. The app allows the students to program an Arduino-compatible development board using the Bluetooth protocol through graphical blocks as Figure 3 shows. The blocks were created using the Google tool Blockly, which is a web engine tool that has been widely employed to teach programming languages from school to university levels and it acts as a parser and translator for the graphical blocks towards the selected programming language [53–55]. Using the blocks, the students can construct algorithms and interact with elements interfaced with the development board such as LCD, IoT devices, sensors, or robots. Additionally, the students can see at any moment the respective code in C language for their algorithm to learn directly the syntax of this language which is typically used in microcontroller or microprocessor courses offered to the students in the engineering curricula. The app was used in two courses (Introduction to Electronics 2020-I and Automation 2021-I) because several students indicated that did not have the availability of a computer, but they had a smartphone. Additionally, one purpose of the methodology was to provide different ways of engagement to learn electronics, programming, and PhyC to the students and the app is suitable for this aim. Then, *EmDroid* was included in these courses to promote ubiquitous learning for the students even in the home spaces. Practitioners and researchers keen on the methodology can find alternatives to this app, for instance, *ArduinoDroid* or *BlocklyDuino* [56,57]. The algorithms created through the graphical blocks in the app are known in computer science as Algorithm Visualizations (AVs). Noone and Mooney [58] have stated that AVs allow students to manipulate the underlying code in a graphical fashion instead that the traditional text-based approach. The AVs have been demonstrated to be an excellent vehicle to foster motivation in students in the early stages of formation at wide educational settings from high school to university courses. Concerning the results and implications of AVs, Hundhausen et al. [59] have conducted a meta-study that shows how AVs influence positively the learning of programming languages based on the analysis of 24 major studies. In the same line, Rößling and Naps [60] depict the impact of AVs on the engagement of students. According to these authors, students can create visualizations to test hypotheses about their algorithms and see their behavior, which contributes both to interchange ideas and debating among peers in a constructivist way. Given these important features for learning, the AVs have been included in the app as a mechanism in which the students can construct and interact with their algorithms graphically. Apart from this app and the Arduino IDE, the software TinkerCAD, and Proteus VSM were employed to simulate the different circuits and the behavior of the development boards with the codes created by the students in textual or graphical mode. Moreover, the app was designed for easy handling by the students to perform the different PhyC tasks.

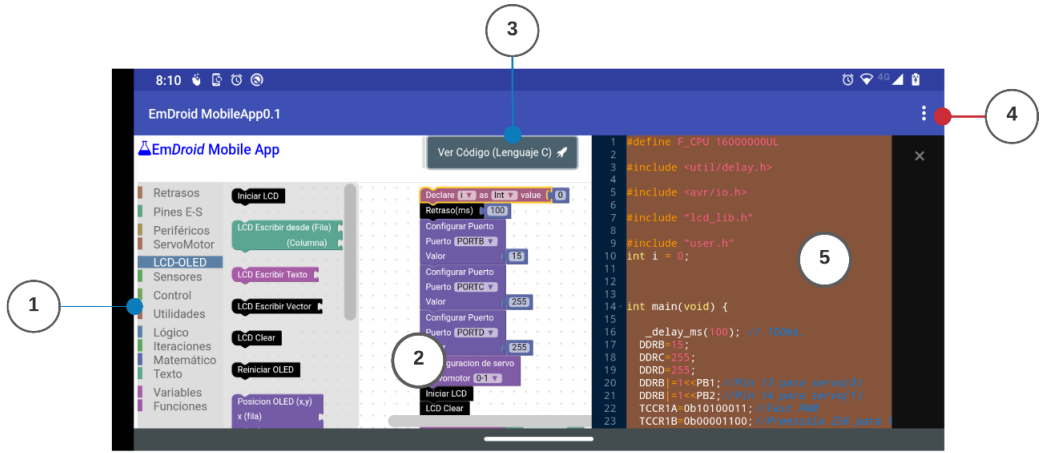

**Figure 3.** *EmDroid* App. 1. Graphical blocks. 2. Working area to construct the algorithms 3. Button to see C language code of the graphical blocks. 4. Main menu. 5. Tab with the equivalent C language for the blocks.

### 2.4 Procedure

The procedure in the courses covers the transitions reported in Figures 4.1-4.3. In these, the different topics, and events concerning the revision of blogs and videos by terms are represented altogether with the exams, and the application of the instruments and techniques to collect data of the students to respond to the RQs proposed in the study. Also, the app EmDroid was incorporated into the classes since the ST in week 7 in the selected courses as well as the activities with PhyC to contrast the academic performance of the students.

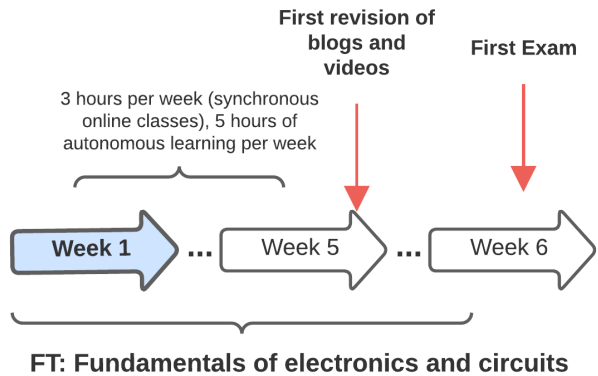

**Figure 4.1.** Procedure adopted for the FT in the courses.

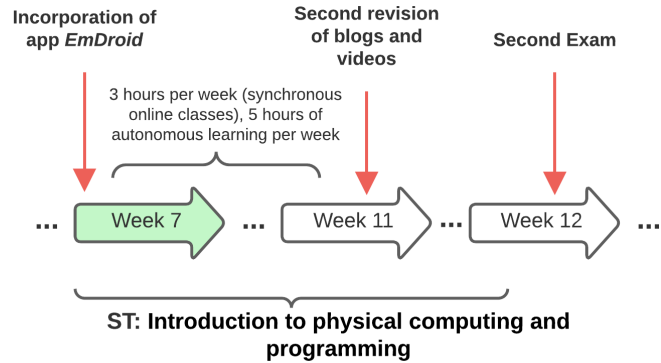

**Figure 4.2.** Procedure adopted for the ST in the courses.

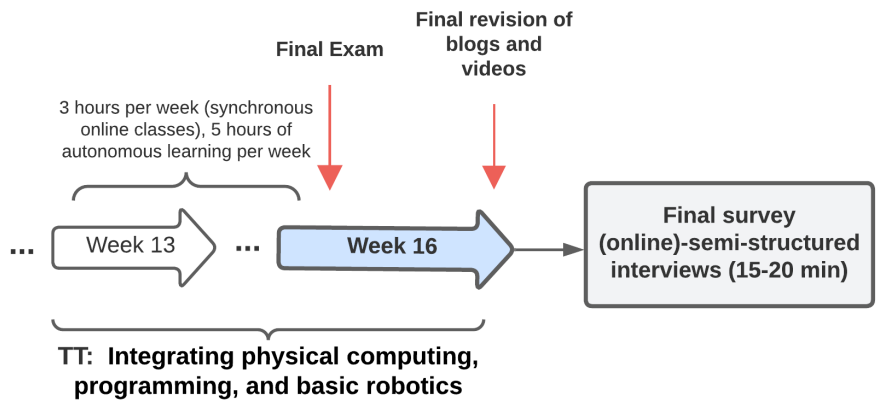

**Figure 4.3.** Procedure adopted for the TT in the courses.

Under the procedure of Figures 4.1-4.3, the students had two sessions of online synchronous classes per week with a total duration of 3 hours. The topics addressed in the classes encompassed fundamentals of circuits, handling of laboratory instruments (multimeter and voltage supply), introduction to physical computing and programming, and integration between PhyC, programming, and basic robotics. Each synchronous online session of 1.5 hours was divided by the instructor into slots of 30 min to address a new topic in a theoretical way, 20 min to review doubts of previous sessions or with the new concepts, and 40 minutes to explain through a live tutorial the way to apply the concepts with the in-home laboratories. Aside from the synchronous classes, each student should study 5 hours per week autonomously. Besides, the blogs and the videos created collaboratively in groups were feedback to the students to improve them before the application of each exam per term (see Figure 4). With the feedback, the students identified and corrected errors both in the theoretical concepts and in their application. In the third term (TT), in week 16, the students delivered the final version of the blogs and videos that were analyzed under the lens of an evaluation rubric with the categories of *organization, content coherence, concept comprehension, concept application, and teamwork*. Finally, the students responded to a final online survey with several closed-ended and open-ended questions. Also, eleven students were selected to respond to a semi-structured interview during a time between 15 min to 20 min. These data were analyzed to respond to each RQ proposed in the study and identify the pertinence of the methodology.

**2.5 Instruments**

It is worth mentioning that due to the number of students in each course and the number of courses per semester a quasi-experiment was not possible to perform. Instead, several sources to collect, analyze, and contrast the data were considered to reduce the validity threats of the study. Through the instruments and their data analysis, it was identified if the methodology was appropriate for both the students and the purposes of the study. To respond to the RQs, firstly, a survey was created with the categories depicted in Table 4. These categories respond to the purposes and learning goals of the study. Table 4 indicates the theoretical or empirical sources for the questions that were reviewed and tailored. The questions of the survey per category and semester can be reviewed in the link available in the supplementary materials in this document. Moreover, the questions were tailored according to the type of learning (online or online with mobile) provided to the students in the courses.

**Table 4**. Description of the categories, the range of the number of questions (N), and the sources for the questions of the surveys.

| Category | Description | N | Source |
|---|---|---|---|
| *Learning and Experimentation* | Measures the perception of the students regarding the in-home laboratories, experimentation, and interaction with the mobile platform for PhyC (*EmDroid*) | 5 | Own construction |
| *Learning and content creation with digital media* | Measures the implications of digital media creation through blogs and videos in the learning and reflection processes of the students. | 4 | Own construction |
| *Motivation, Anxiety, and Self-efficacy* | Measures the motivation, self-efficacy, and reduction of anxiety in the courses with the methodology | 2-4 | Engineering self-efficacy measurement [61], the situational motivation scale (SIMS) [62], and own construction |
| *Teacher Influence and Feedback* | Estimates the influence of the teacher and his feedback in the learning process | 1-2 | Teacher presence in the community of inquiry (CoI) [63], own construction |

The number of questions in the survey varied in the courses between 12 closed-ended questions for 2020-I, 14 for 2020-II, and 15 for the period 2021-I, and three open-ended questions for all surveys. The increase in the number of questions depended on the improvement of the instruments in each course with the inclusion of other variables such as self-efficacy and reduction of nervousness following the evolution of the educational context produced by the COVID-19 pandemic, or the modality combining online, or mobile learning offered to the students. Each closed-ended question in the categories had a Likert scale in the range (1) Strongly disagree; (2) Disagree; (3) Agree; (4) Strongly agree. To measure the internal reliability of the survey (instrument) per category, it was employed the McDonald's omega coefficient ($\omega$) with the values shown in Table 5 according to the course and a confidence interval of 95%. The McDonald's omega coefficient is a well-known statistical alternative to Cronbach's Alpha ($\alpha$) based on factorial loads and is more suitable when the number of options of the survey is less than the traditional 5 points Likert scale as is the case of the survey in this study [64,65]. Like ($\alpha$), a value over 0.7 demonstrates good reliability of the instrument in its categories. N=53 (94.64%) students answered the survey anonymously in the courses. In some cases, due to the number of questions and respondents was not possible to calculate the ($\omega$) coefficient.

**Table 5.** Internal consistency of the survey by category, number of respondents (N), and course through McDonalds' omega coefficient ($\omega$). I, II indicate the semester of the year.

| Category/Dimension | Period | N | $\omega$ |
|---|---|---|---|

| | 2020-I | 14 | .815 |
|---|---|---|---|
| *Learning and Experimentation* | 2020-II | 25 | .77 |
| | 2021-I | 14 | .98 |
| *Learning and content creation with digital media* | 2020-I | 14 | .851 |
| | 2020-II | 25 | .825 |
| | 2021-I | 14 | .851 |
| *Motivation, Anxiety, and Self-efficacy* | 2020-I | 14 | - |
| | 2020-II | 25 | .769 |
| | 2021-I | 14 | .644 |
| *Teacher support and feedback* | 2020-I | 14 | - |
| | 2020-II | 25 | - |
| | 2021-I | 14 | - |

In this case, .644 ≤ω ≤ .98, which validates the internal consistency of the instrument between acceptable and excellent per category. Secondly, to find complementary perceptions of the students regarding the methodology, a set of semi-structured interviews were developed with a duration between 15-20 mins. The interviews addressed additional questions regarding the methodology, the mobile platform used in the courses, and the initiative with the in-home laboratories for learning and motivation. The interviews were transcribed in Microsoft Word format for further analysis. Thirdly, to evaluate the blogs and videos of the students, it was constructed an evaluation rubric with the criteria shown in Table 6, and an evaluation range per criteria between 1 (very deficient) to 5 (excellent). The rubric is available in the section of supplementary materials. Finally, to respond to RQ3, it was registered the grades of the students in each term to see their progression in the courses with mobile and online learning integrated vs. the one with only online learning. These four data sources allowed us to respond to the RQs and support the findings of the study.

**Table 6.** Rubric categories and criteria.

| **Rubric category** | **Criteria** |
|---|---|
| *1. Organization* | The organization of the blog is easily readable, without errors in the access of the documents or information resources, and the entries are in order by topics or concepts. |
| *2. Relevance and coherence in the contents* | The contents presented in the blog are coherent, synthesized, and related to the topics and concepts of the course. |
| *3. Concept Comprehension* | The contents presented in the blog and in the videos evidence an adequate comprehension of the topics and concepts of the course. |
| *4. Concept and knowledge application* | The videos and resources of the blog evidence an application of the learned concepts through the in-home laboratories and/or the mobile app. |
| *5. Teamwork* | The videos and resources in the blog demonstrate teamwork. The contents, videos, and laboratories were made collaboratively by the members of the group. |

**2.6 Analysis**

To analyze the data and respond to the RQs, it was adopted a mixed approach through embedded design that is employed in engineering education studies [66]. In this, the quantitative and qualitative data collected with the mentioned instruments are analyzed parallelly to find the outcomes of the study and contrast them with the theoretical concepts or constructs. For the analysis of quantitative data of the surveys and students' grades, the software IBM SPSS v.27 was used, while *NVIVO* v.12 was employed for the qualitative data of the surveys and the semi-structured interviews. To respond to the RQ1, it was analyzed the descriptive statistics in terms of mean (*M*) and standard deviation (*SD*) of each category in the survey, as well as the average of the score in the criteria of the rubric per course. To respond to RQ2, it was performed some Pearson correlation matrices with the categories of the survey to find the relationships between the constructs of learning, experimentation, self-efficacy, learning with the creation of digital media, and teacher feedback and support. Also, the semi-structured interviews and the comments of the students in the surveys were analyzed under the lens of thematic analysis [67]. This technique allows for interpreting the manifest content of the communications making a rigorous analysis and considering aspects such as completeness, homogeneity, relevance, and exclusiveness of the information [67]. Through this technique, the common themes in the interviews were identified, which nourished the findings of this work and revealed additional implications of the methodology. A content-driven approach was adopted to perform this analysis, allowing the themes that emerged from the comments of the students. At last, to respond to RQ3 was analyzed the students' grades in the different courses by term. Due to the nature of these data, the non-parametric Wilcoxon signed-rank test was performed to find statistical significance in the grades with the methodology between the FT and TT. The results and discussion will be presented according to the proposed Research Questions (RQs).

## 3. Results and discussion

**3.1** *RQ1. What are the students' perceptions regarding learning, experimentation and motivation with the in-home labs and the creation of blogs and videos in the online courses?*

To answer this question, Table 7 shows the descriptive statistics of each category of the survey applied to the students. In the category *Learning and Experimentation*, the questions Q1-Q5 sought to identify if the in-home laboratories, the methodology, and the app *EmDroid* had an impact on the learning of the students. In general terms, the students had a good perception of strategies adopted in the methodology for their learning with better results for the course 2020-II ($M = 3.87$, $SD = .324$), where only online learning modality was used (see Table 7 and Figure 5). While in this course, 100% of the students had answers between agree and strongly agree, the courses in which the app *EmDroid* was deployed had a different distribution. In special, one student in the course 2020-I manifested that already knew to program and he expected that the app had another form of programming with a textual-based mode in C language which for him was more interesting. In other cases, for instance, in the course 2021-I, some students had problems with the app and the experimentation with the in-home laboratories because they had iOS phones and not Android OS equivalents.

These issues regarding accessibility, affordability, or usability of apps that employ mobile learning have been documented in the references [68–70] and they influence the learning and motivation of the students in the courses. Students had a positive perception of mobile systems when they can control their learning by themselves [68]. To solve this accessibility problem, several tablets with the app *EmDroid* were facilitated to the students. This fact could have had an impact on the perception of the students in the category and it has been mentioned by the literature as "facilitating conditions". This factor describes the guidance and technical assistance to facilitate engagement with learning systems [71].

Thus, despite the problems mentioned, most of the students in the courses with the *EmDroid* app ($n=27$, 96.43%) agreed or strongly agreed with Q1 "*EmDroid helped me to understand the interaction between hardware and software in an electronic system*". For Q2 "*EmDroid allowed me to understand the programming steps (algorithm) used to interact with the hardware devices*" and Q3 "*EmDroid allowed me to understand the relationship between a graphical block, its function, and its representation in C language*", ($n=25$, 89.29%) of the students agreed or strongly agreed. Similarly, ($n=26$, 92.86%) agreed or strongly agreed regarding Q4 "*EmDroid and the course methodology aided me to perform the different laboratories in the home spaces*", and Q5 "*The developed methodology allowed me to understand the electronics or electrical concepts such as voltage, current, or power, and their application*".

**Table 7.** Descriptive statistics for the categories in the survey according to the course. I, II is the semester of the year. Online Learning (OL), Mobile Learning (ML).

| Category/Dimension | Period | Learning Modality | M | SD |
|---|---|---|---|---|

| 1. Learning and Experimentation | 2020-I | OL+ML | 3.43 | .665 |
|---|---|---|---|---|
| | 2020-II | OL | 3.87 | .324 |
| | 2021-I | OL+ML | 3.5 | .833 |
| 2. Learning and content creation with digital media | 2020-I | OL+ML | 3.51 | .642 |
| | 2020-II | OL | 3.75 | .427 |
| | 2021-I | OL+ML | 3.73 | .527 |
| 3. Motivation, Anxiety, and Self-efficacy | 2020-I | OL+ML | 3.48 | .681 |
| | 2020-II | OL | 3.69 | .48 |
| | 2021-I | OL+ML | 3.5 | .524 |
| 4. Teacher support and feedback | 2020-I | OL+ML | 3.57 | .513 |
| | 2020-II | OL | 3.87 | .344 |
| | 2021-I | OL+ML | 3.78 | .416 |

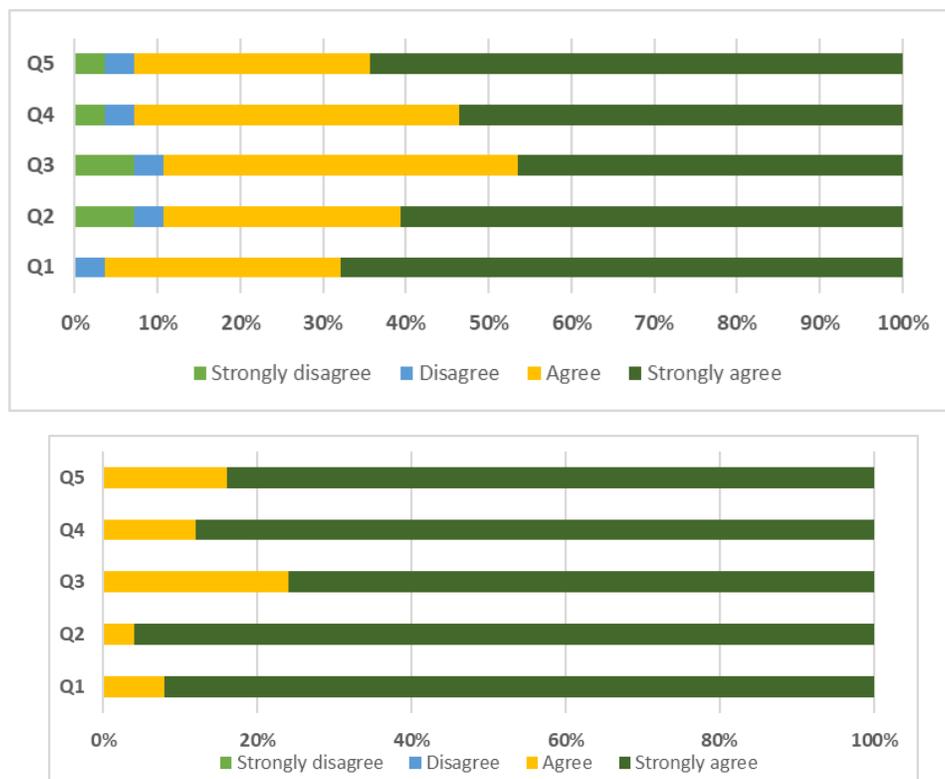

**Figure 5.** Perception of the students by question in the category of Learning and Experimentation. *Top*: results for the periods 2020-I and 2021-I (Online learning + mobile learning, *n*=28), *Bottom*: results for the period 2020-II (Online learning, *n*=25).

Similar results were identified for the course 2020-II with only online learning. As described this course employed the Arduino platform for the PhyC activities. In this way, (*n*=25, 100%) agreed or strongly agreed for Q1 "*The in-home laboratories allowed me to understand the electronics concepts in the courses*" (see Figure 5). A comparable distribution is repeated for the questions Q2 "*The in-home laboratories allowed me to apply the concepts learned through experimentation*", Q3 "*The proposed problems along the course allowed me to learn with the in-home laboratories*", Q4 "*The in-home laboratories helped that the class was more dynamic and interesting*", and Q5 "*I consider that the in-home laboratories are a good strategy that fosters my learning*".

Regarding the second category *Learning and content creation with digital media*, students made a total of 24 blogs with the different videos in the courses. Figure 6 illustrates some examples of blogs and videos made by the students. The results in Table 7 suggest a good acceptance of the students concerning the creation of digital media. The better results were obtained for the courses 2020-II (*M*=3.75, *SD*=.427) and 2021-I (*M*=3.73, *SD*= .527). This trend was due to the evolution of the methodology with digital media in the courses and the context of COVID-19 in the first semester of 2020. In the middle of this semester, many HEIs restricted the face-to-face classes and the students and teachers migrated quickly to an ERT format which was difficult for many of them. Some students in the course 2020-I indicated difficulties in the adaptation to the tool Google sites used for many of them to construct their blogs and the way to organize and distribute the different contents in the blogs. Therefore, for the next semesters, the instructor destined some class sessions to show different examples of blogs and clearly define the requisites and evaluation criteria for them. These comments of the students are congruent with the categories evaluated with lower scores in the blogs according to the rubric criteria in Table 6. The overall descriptive statistics for the rubric criteria in each course evidence a trend towards a good performance for (2020-I: *M*=4.16, *SD*=.4373), (2020-II: *M*=4.42, *SD*=.36), and excellent for (2021-I: *M*=4.562, *SD*=.392). However, students in the courses of Automation scored less in the categories of *organization and relevance in the contents,* whereas the students in Introduction to Electronics did it in *concept comprehension and application* (see Table 7). Readers can find some URLs with examples of the blogs and videos in the supplementary materials.

**Table 7**. Overall scores obtained by rubric category for the blogs and videos evaluated in the courses.

| Rubric category | Course | | | | | |
| --- | --- | --- | --- | --- | --- | --- |
| | 2020-I | | 2020-II | | 2021-I | |
| | M | SD | M | SD | M | SD |
| *1. Organization* | 4.2 | .524 | 4.39 | .404 | 4.28 | .311 |
| *2. Relevance and coherence in the contents* | 4.28 | .5 | 4.41 | .375 | 4.48 | .435 |
| *3. Concept Comprehension* | 4.1 | .56 | 4.51 | .36 | 4.71 | .5 |
| *4. Concept and knowledge application* | 4.1 | .41 | 4.6 | .31 | 4.75 | .384 |
| *5. Teamwork* | 4.2 | .25 | 4.19 | .33 | 4.6 | .33 |

In particular, these scores in the categories of *organization and relevance* are related to a concept known as "content curation", which is defined as the skill to investigate, find, filter, organize, edit, and share the best and most relevant information about a topic in an online digital collection [72]. Under the plethora of information, students must investigate, select, organize, and share the information in their blogs, which in this case was difficult for some students in the courses. This result has been evidenced by other studies in the field of engineering, personal learning environments (PLEs), and content curation that illustrate how students could struggle with the skill of *search* and *sense-making*, that is, the skill to create content, adding value, and personalized them [73–75]. Similarly, the previous data suggest that the students of the first semester (Introduction to Electronics-2020-I) are more focused on the content and organization in their blogs and a little less on understanding and applying the concepts entailed.

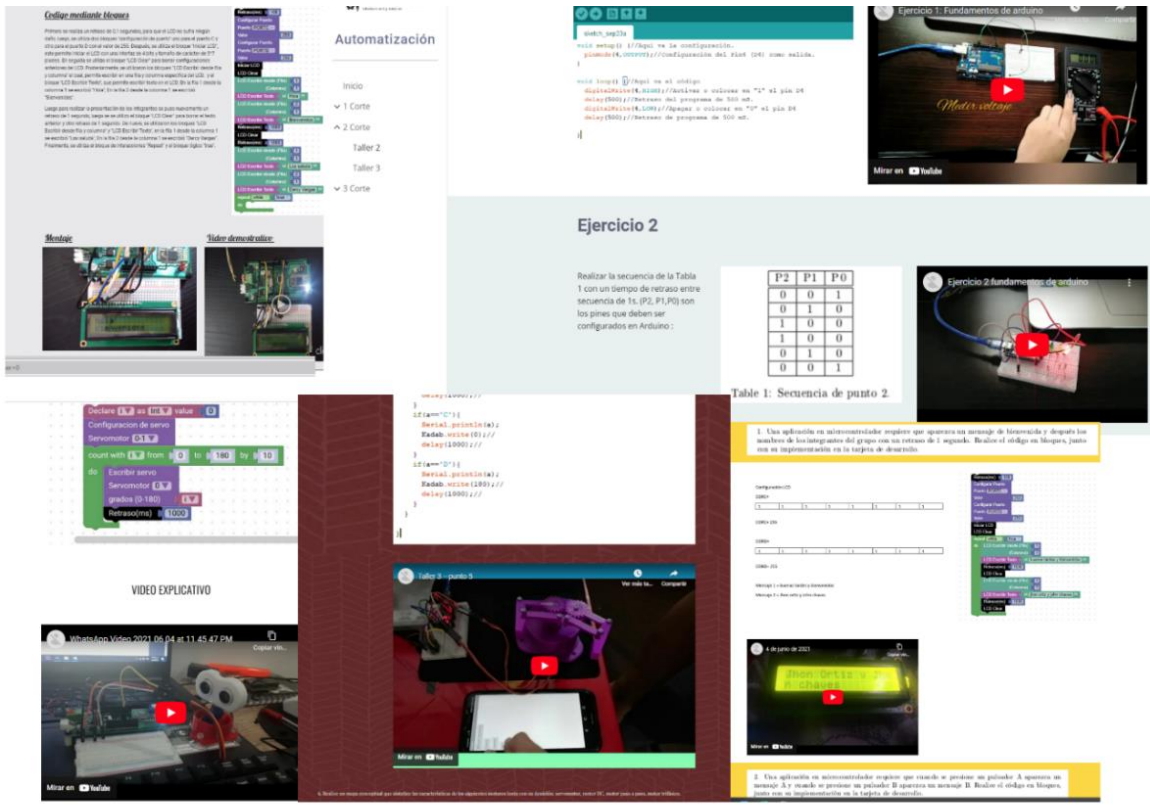

**Figure 6.** Examples of blogs and videos made by the students in the courses.

In contrast, in the courses of Automation (eighth semester of industrial engineering-2020-II, 2021-I), the students had an excellent performance in the categories of concept comprehension and application. Further incorporation of strategies that level all aspects of the rubric categories towards an excellent performance is needed in the courses, integrating efficient handling of ICT. Another consideration is the importance to reinforce teamwork not only with the availability of online resources but promoting more interaction between the students in a group through meetings, forums, personalized tutoring, etc. Aside from the previous aspects, Figure 7 shows the perception of the students with the surveys regarding the inclusion of digital media with blogs and videos for their learning. This category had the same questions in the surveys for all courses.

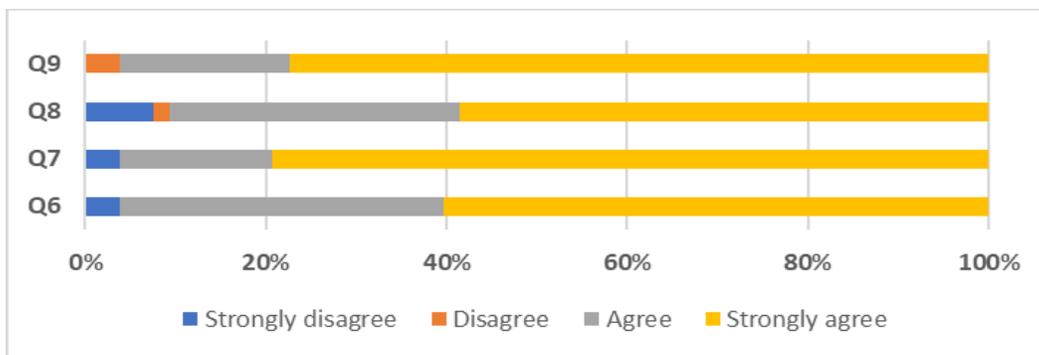

**Figure 7.** Perception of the students by question in the category of *Learning and content creation with digital media (n=53)*.

For Q6 "*through the creation of the blog, I reinforced the topics learned in the course*", most of the students (*n*=51, 96.23%) agreed or strongly agreed. Similar results are achieved for Q7 "*I consider that the blog was a good strategy to follow up my learning process*", Q8 "*The creation of the blog and videos allowed me to improve my skills in handling ICT*", and Q9 "*Through the creation of the blog and videos, I worked collaboratively with my peers*".

One central point in the methodology apart from the learning and experimentation was the increase in motivation, self-efficacy, and the reduction of anxiety patterns experienced by the students. As it was mentioned, this was achieved through an extension in the time to deliver exams, the inclusion of ill-structured problems for PhyC activities that required the usage of the hardware materials in the in-home laboratories, software, the creation of blogs and videos, and the teacher support. Especially, ill-structured problems have demonstrated to foster creativity and high-order thinking skills in students in online engineering education [76]. It is worth mentioning that self-efficacy refers to the personal beliefs about possessing capabilities to cope with determined situations and get the desired results [77,78]. Self-efficacy is correlated to students' dropout and academic performance [43,79]. Indeed, of 56 students, only 1 student (1.78%) dropouts the course of Introduction to Electronics by personal reasons.

Students in the category of *Motivation, Anxiety, and Self-efficacy* in the courses of Automation had the perceptions depicted in Figure 8. Here, for instance, for Q10 "*The in-home laboratories and the proposed methodology motivated me to learn in the course*" (*n*=38, 97.44%) of students agreed or strongly agreed, for Q11 "*Comparing my abilities and/or skills at the beginning of the course, I feel more prepared and confident to face the problem-solving process*", and Q12 "*The methodology helped me reduce my nervousness towards the delivery of exams and activities in the course*", the (*n*=39,100%) of students agreed or strongly agreed in these. A comparable percentage distribution in Q10 is evidenced for Q13 "*I consider that my performance in the course was good*" (see Figure 8).

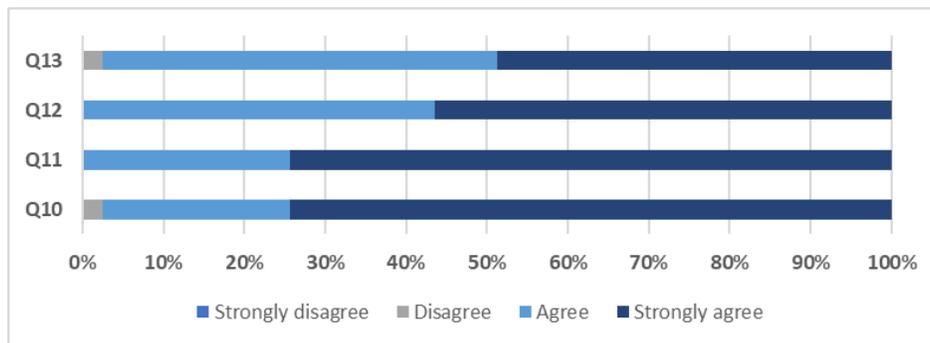

**Figure 8**. Perceptions of the students for the Automation courses (2020-II, 2021-I, *n*=39) in the category Motivation, Anxiety, and Self-efficacy.

For the course Introduction to Electronics, the results are like the automation course for Q10 "*the topics and activities were of my interest and motivate me to learn*", and Q11 "*the use of the in-home laboratories and the proposed methodology motivated me to learn in the course*". (*n*=14, 100%) of the students agreed or strongly agreed for Q10, and (*n*=12, 85.71%) for Q11.

A factor that has a relationship with motivation, self-efficacy, and reduction of anxiety according to the results was the teacher support and feedback that was evaluated by the students in the surveys. For instance, in the surveys, all students indicated that the constructive feedback provided was beneficial for their learning process. For the course of Automation (2021-I), (*n*=14, 100%) of the students agreed or strongly agreed in Q15 "*I consider that the open attitude of the teacher allowed me to have more confidence to address my concerns and doubts*". Previous results in the *Teacher support and feedback category* are aligned with what is expected in the creation of significant learning experiences of the ICDM [33] and with the engagement of students in online learning settings [80]. The teacher not only acts as a transmitter of knowledge and information but acts as a scaffolder and facilitator of the learning process which is a clear differentiator of AL. As an observation, the adoption of an open and constructive standpoint of learning instead of a rigid one can contribute to the students to feel more confident and motivated in classrooms and learn more about PhyC, programming, and electronics.

**3.2** *RQ2. What factors influence active learning, experimentation, and motivation in the courses combining online and mobile learning?*

Tables 8 and 9 show the Pearson correlation matrices for the survey categories. The Pearson correlation was performed after the normal distribution of data was tested through the Kolmogorov-Smirnov test. The matrices were divided by gender into the courses. This division was due to the analysis of all data did not yield significative correlations but with the segmentation, there are differences between men and women students in the courses. Stronger correlations are highlighted in gray. For instance, the teacher support and feedback (TSF) have strong correlations with learning and experimentation (LE) ($r(16)=.935$, $p<.001$), learning with digital media (student-created videos) (LDM) ($r(16)=.819$, $p<.001$), and the motivation and self-efficacy ($r(16)=.844$, $p<.001$) in the case of women students. Otherwise happens for men students where the correlations in Table 9 are weak for the category TSF. Besides, motivation, reduction of anxiety, and self-efficacy have strong correlations with LE with ($r(16)=.935$, $p<.001$) for women, ($r(39)=.878$, $p<.001$) for men, and with LDM with ($r(16)=.871$, $p<.001$) for women, ($r(39)=.618$, $p<.001$) for men students. Weak or even negative correlations were identified for age concerning all categories in the surveys. This suggests that age does not play an important role in the methodology which had good acceptance among all students.

**Table 8**. Pearson correlation matrix for women students. LE: Learning and Experimentation, LDM: Learning and content creation with digital media, MAS: Motivation, Anxiety, Self-Efficacy, TFS: Teacher Feedback and Support. ** $p<.001$. $n=16$.

| Category | LE | LDM | MAS | TFS | Age |
|---|---|---|---|---|---|
| 1. LE | 1 | | | | |
| 2. LDM | .836** | 1 | | | |
| 3. MAS | .834** | .871** | 1 | | |
| 4. TSF | .935** | .819** | .844** | 1 | |
| 5. Age | -.036 | -.142 | -.247 | .051 | 1 |

**Table 9**. Pearson correlation matrix for men students. LE: Learning and Experimentation, LDM: Learning and content creation with digital media, MAS: Motivation, Anxiety, Self-Efficacy, TFS: Teacher Feedback and Support. ** $p<.001$. $n=39$.

| Category | LE | LDM | MAS | TFS | Age |
|---|---|---|---|---|---|
| 1. LE | 1 | | | | |
| 2. LDM | .673** | 1 | | | |
| 3. MAS | .878** | .618** | 1 | | |
| 4. TSF | .232 | .255 | .088 | 1 | |
| 5. Age | .116 | -.204 | -.147 | .024 | 1 |

Along with the previous data, thematic analysis ratifies the mentioned factors with the codebook and commentaries examples shown in Tables 10 and 11. These codes emerged from the interpretation of the comments of the students based on their frequency (*f*) in the transcriptions of the semi-structured interviews and the comments in the open-ended questions of the surveys. The first code evidenced is the *Teacher characteristics that support learning*. Several frameworks widely spread and adopted in online learning settings such as the Community of Inquiry (CoI) [63,81,82], integrating this code as *Teacher Presence*. Students mentioned a set of teacher characteristics that allowed them to successfully learn such as openness, communication, feedback, willingness, and recursion. From there, willingness is the theme with the most frequency in the comments. Willingness in the sense that the teacher provided a learning space to the students, responding to their learning needs and doubts in the course. Students mentioned in the same theme, the quality of the explanations in classes, the feedback, and the personalized tutoring as remarkable aspects of the teacher. Similarly, openness and communication refer that the teacher had an open dialog that allowed the closeness with the students to make their queries and express them without coercion. The recursion is associated with how the teacher found effective options to teach the different concepts and topics in the courses.

The second code is the *features of the mobile learning app* that were manifested by the students in the courses that employed the app *EmDroid*. Thus, affordability is the theme with the most frequency. In this, students mentioned diverse factors such as the easiness to program and making physical computing activities, the facility to establish a Bluetooth pairing with the Arduino-compatible development board, and the distribution of the graphical blocks as aspects that contribute to the affordability of the application. For most of the students, block-based programming was a novel and good feature of the app because it brings closer the programming and PhyC for novice programmers without excessive effort. This is particularly useful for engineering students approaching the concepts in electronics, physical computing, and basic robotics in introductory courses. In addition, the possibility to interact with hardware devices and experiment with the app through the construction of algorithms with the graphical blocks was another theme exposed in the comments. Several students mentioned that the possibility to make PhyC was a way in which they can contrast their block-based programs with the C language and find an application for them in daily life. This leads to the final theme in this code, which is the *support for motivation* in which the students manifested that the features of the app allowed to motivate them even for students who do not find programming easy. With these features, students indicated their intention to use the application in other related courses. Some of the previous aspects especially the affordability and the intention to use have been explored in prior studies related to models of acceptance of the technology for mobile learning such as the Technology Acceptance Model (TAM) or The Unified Theory of Acceptance and Use of Technology (UTAUT)[71]. Nonetheless, a student who already knew how to program indicated that the app needs the inclusion of a text-based mode which can add more functionality to the app. This can be an improvement for further work.

The third code was the *Learning and Experimentation with hands-on activities* mentioned by most of the students in the courses. The hands-on activities allowed to put into practice the learning of the concepts, foster problem-solving skills in the students and gave dynamism to the classes. Indeed, students indicated that this was a good strategy that promoted motivation in the worst period of the COVID-19 pandemic. This was possible thanks to the in-home laboratories with the hardware components sent to the home of each student. This was mentioned in the fourth code in the thematic analysis as the *Facilitation of conditions to learn and experiment*. This code denotes the conditions in technology infrastructure with LMS and the usage of smartphones or PCs to attend to the synchronous classes, the applications to record and edit the videos and make the blogs, the educational features of the mobile app *EmDroid* mentioned, the experimentation with the in-home laboratories through the possibility to access hardware components, the teaching resources such as explanation videos, lectures, or live laboratories in the classes, and the teaching methodology mentioned in the previous results that allowed and enhanced the learning process of the students. For instance, without the hardware components in the kits, the hands-on activities would have been difficult to fulfill in the engineering courses. Likewise, without the features of the mobile app, learning programming, PhyC, and basic robotics would have difficult for most students. Then, the facilitation of conditions such as technical and educational seems to play an essential role in online and mobile learning settings in engineering education.

The fifth code was the *Experiences with the creation of digital media* with blogs and student-created videos. These types of digital media constructed by the students had three aims. The first one was to evaluate the skills developed by the students in electronics, PhyC, and programming in a different way that the traditional use of closed-ended questions in exams. Although was employed this last type of exam in the courses, the questions of these were ill-structured problems, and an additional time to deliver these exams with video creation was provided to the students. This type of student-created video is denominated a performance-orientated project in which the students exhibit different performance skills [48]. A second aim was that the students reflected on their process of learning and if they understood and applied the concepts adequately. In this way, although the methodology did not pursue directly metacognitive skills according to the learning outcomes in Table 2 in the function of Marzano's taxonomy, the blogs and in special the student-created videos encourage these skills. Students indicated that the videos were a good strategy to identify the comprehension of the topics in the courses. In addition, the blogs and videos foster in-depth investigation of additional topics and promote the development of ICT skills and teamwork (see comments in Table 11). Nevertheless, as the Pearson correlation matrix for women students illustrates, the learning and reflection with these types of resources go accompanied by teacher support and feedback. The last aim was to provide a method to follow up the learning process of the students. In accordance with the comments, this was possible through the creation of blogs and videos and was manifested by the students. As aspects to improve, the students indicated extending class time and shortening points for assignments and midterms which will consider in future implementations of the methodology in other courses of the engineering curricula.

**3.3** *RQ3. What are the educational implications of the methodology for the academic performance of the students?*

Figure 9 illustrates the violin and box plots for the students' grades in the courses ($n=56$). In this case, the grades are in the range 0-5. For the 2020-I course (FT: $M=3.8$, $SD=.71$ - TT: $M=4.36$, $SD=.4$), for 2020-II (FT: $M=4.4$, $SD=.36$ - TT: $M=4.55$, $SD.32$), and for 2021-I (FT: $M=4.12$, $SD=.65$ -TT: $M=4.56$, $SD=.41$). In general, students tend to score lower in the FT with the topics of fundamentals of electric circuits that according to the comments in the interviews and surveys were difficult for several students. In this term was used the in-home laboratories with an important load of theoretical concepts starting with Ohm's law, serial, parallel, and mixed circuits. Notice that the improvements in the grades started from the incorporation of PhyC activities in ST and continue for the TT where were entailed the integration of PhyC activities, basic robotics, and programming. Hence, the results suggest that the incorporation of PhyC activities in the courses had a positive impact on the academic performance of the students. In addition, although the results in the perceptions in RQ1 indicated some problems in the accessibility with the app *EmDroid*, better improvements in the grades were achieved in the courses that incorporated it (see courses 2020-I and 2021-I in Figure 9). This suggests that the integration between online and mobile learning leads to better results for the academic performance of the students.

**Table 10.** Codebook for the thematic analysis of semi-structured interviews and comments in the surveys. *f*: commentary frequency

| Code | Theme | *f* | Commentary examples |
|---|---|---|---|
| 1. Teacher characteristics that support learning | -Willingness | 18 | "*It seems to me that as much as possible the teacher tried to answer all our questions*"<br>"*An excellent teacher, really the way of explaining is very cool, he was always there for his students, he always gave the space to help*" |
| | -Feedback | 7 | "*The immediate feedback received and the possibility of continuous improvement during and outside the classes*"<br>"*The support provided by the teacher and the feedback given after each delivery*" |
| | -Openness, communication | 4 | "*It is a very interesting course, I learned a lot about the topics from the teacher, who was totally open to doubts and explained patiently in each class*"<br>"*Good communication, respect in class and the appropriate methodology which allowed a better understanding of the topics*" |
| | -Recursion | 4 | "*The professor searched for ways to project us his screen, making the circuit assemblies, and using all the additional applications to make it easier for us to understand the topics*"<br>"*Thanks to the videos made by the instructor of the different topics, I had a good understanding of the topics*" |
| 2. Features of the mobile learning app | -Affordability | 16 | "*The ease with which it is used, it is not too complex to understand and that helps students*"<br>"*Easy to understand the programming from this application (EmDroid), since I did not understand much of the basic programming subject viewed along the semester, and I did not know how to apply this knowledge in everyday life*" |
| | -Block-based programming | 13 | "*The block programming was different from what is commonly done. It was easy to use and allows me to understand the use of the different blocks and, thanks to this, it was a platform that made it easier for me to learn*"<br>"*The most interesting thing was the creation of code with blocks since you don't need to be an expert in programming to be able to handle it*" |
| | -Experimentation with hardware | 6 | "*It is a very complete platform for the use of LCD and OLED displays, temperature sensors and mainly the Atmega328p microcontroller*"<br>"*Interacting with the hardware elements did allow me to better understand the programming, since, although the platform is very complete, the process is complemented with the practice of these elements. The interaction allows us to understand what we are reproducing in the block codes*" |
| | -Support for motivation | 5 | " *Yes, I would use the platform again. I thought it was a good learning strategy that could be used in both modalities (face-to-face and online)*"<br>"*Programming, I emphasize a lot, does not come easy to me, I have never really liked it, so the block programming of the EmDroid platform did increase my interest towards the subject*" |

**Table 11**. Table 10 continued.

| Code | Theme | f | Commentary examples |
|---|---|---|---|
| 3. Learning and Experimentation | -Hands-on tasks | 22 | "*The methodology is a good initiative that allowed us to learn from home given the contingencies before COVID-19*"<br>"*Putting into practice the knowledge acquired in the Automation class through the workshops and other activities greatly facilitated the understanding of the topics, making each class a pleasant opportunity to learn new concepts and topics*" |
| 4. Facilitation of conditions to learn and experiment | - | 7 | "*Taking into account the situation and that in my opinion virtuality hinders learning, I consider that having adequate work elements (kits), and having clear guidelines for the realization of each workshop, allows for a pleasant learning experience and exploration of possibilities*"<br>"*One thing I emphasized was to supply the students with materials for the development of the proposed tasks and laboratories*"<br>"*The EmDroid tool is a great strategy that allows us to have the possibility to experiment from home.*" |
| 5. Experiences with the creation of digital media | -Deepening of concepts and topics | 6 | "*The realization of the blog helped me to do more research on the subject to go more in-depth*"<br>"*With the blog, I showed what has been learned and investigated other aspects in order to gain more knowledge*" |
| | -Awareness of learning skills | 5 | "*The blog is an educational strategy that allowed us to improve and demonstrate our skills*"<br>"*It seems to me that making an explanatory video demonstrates whether the necessary knowledge was really acquired*"<br>"*In the creation of the blog, the appropriation of the topics of the subject was more evident, which helped me to advance and have some clearer concepts*" |
| | -Follow up | 2 | "*The blog allows an easy grouping of information, easy support and follow up of the process carried out*"<br>"*The blog allows personal tracking of knowledge progress*" |
| | -Teamwork | 3 | "*With the blog and the videos, I worked collaboratively with my peers*" |

Likewise, Figure 10 and Figure 11 show the violin and box plots for the academic performance by gender in the courses 2020-II and 2021-I in Automation. Nevertheless, it is important to clarify that the primary objective of the methodology was not to reduce the gender gap in programming skills but to provide the same learning opportunities for all students. The incorporation of PhyC activities tends to level the academic grades of women and men students towards the range of 4.5-5.0 (excellent), and in the case of the course of 2020-II, the female students had scores above their male peers. Nonetheless, an observation that can lead to this result is that women students used the personalized tutoring services more often than their male peers, solving their doubts with the different concepts and hands-on activities. Besides, the usage of the graphical blocks in *EmDroid* during the Automation course of 2021-I and Arduino in Automation 2020-II altogether with the teacher support could be catalysts of the attitudes shown by women students regarding the learning of programming and PhyC. Also, the descriptive statistics for the course 2020-II for women students in the FT and TT (FT: *M*= 4.425, *SD*= .36 - TT: *M*=4.56, *SD*= .336) and for 2021-I (FT: *M*= 4.425, *SD*=.36 - TT: *M*=4.56, *SD*= .336) reaffirm the previous outcome and observation. These results are aligned with prior studies concerning the closing of the gender gap in abilities and confidence in programming because in these studies men students have better outcomes and grades in programming courses than their women peers [83]. By the same token, the courses involving robotics and programming are preferable to women students and help to produce better educational outcomes for them [83,84].

To identify statistical significance in the previous results between the FT and TT, a set of inferential tests were performed. Here, the null hypothesis is that there is no statistical difference between the grades in the FT and TT depicted in Figure 9, this last term with the incorporation of PhyC activities with basic robotics. The Wilcoxon signed-rank test for the 2020-I course showed a statistically significant between FT (*mean rank*=3) and TT (*mean rank*=7.2) with *T*=6.0, $z = -2.594$ (corrected for ties), N-ties=12, *p*=.009. Comparably for the 2020-II course, FT (*mean rank*=9.25) and TT (*mean rank*=12.97) with *T*=55.5, *z*=-2.537 (corrected for ties), N-ties=23, *p*=.011. Finally, for the course 2021-I, FT (*mean rank*=4.88) and TT (*mean rank*=9.14) with *T*=19.5, *z*=-2.312 (corrected for ties), N-ties=15, *p*=.021. In the previous results, *p*<.05, which demonstrates statistical significance in the courses for the FT vs. the TT.

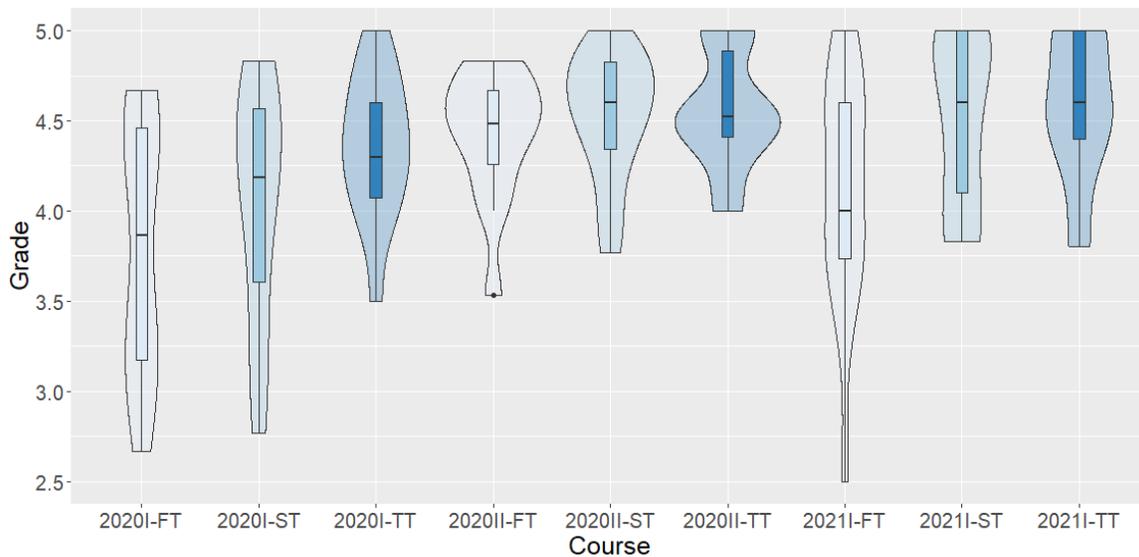

**Figure 9.** Violin and box plots for the students' grades in the courses by term. FT: first term, ST: second term, TT: third term.

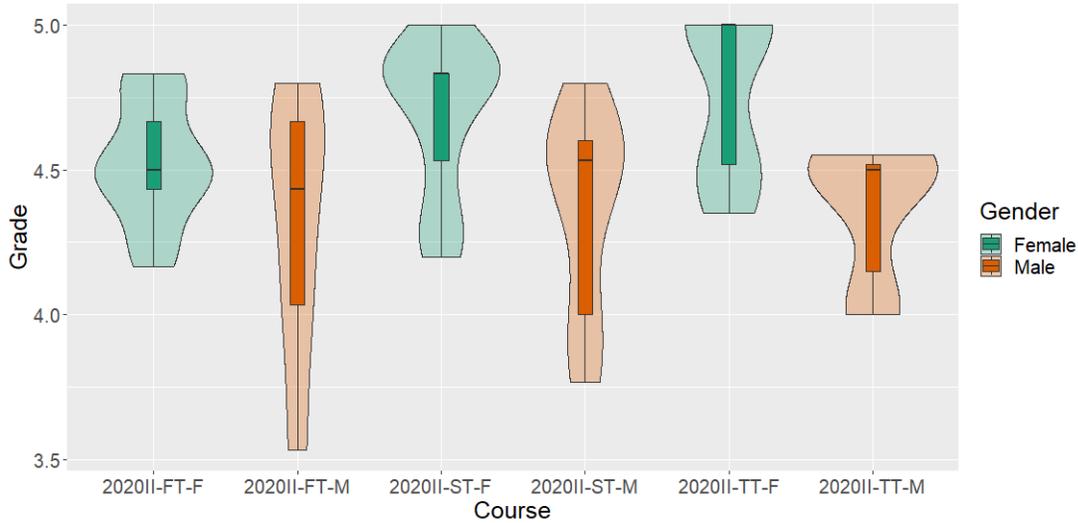

**Figure 10**. Gender comparison of the students' grades in the Automation (2020-II). M= male, F: female. FT: first term, ST: second term, TT: third term.

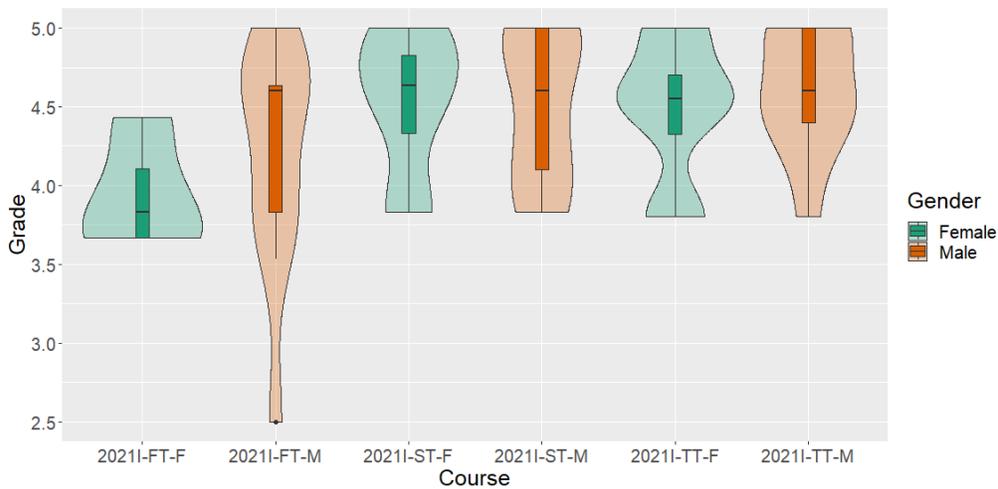

**Figure 11**. Gender comparison of the student's grades in the Automation (2021-I). M= male, F: female. FT: first term, ST: second term, TT: third term.

Therefore, this indicates that PhyC activities, the in-home laboratories with hands-on activities to promote experimentation, and the student-created videos have a positive impact on academic performance. In addition, the mean ranks of the previous tests allow to ratify that there is a better improvement in academic performance in the courses that combine mobile and online learning vs. the one with only online learning. Nonetheless, this was possible due to the features of the mobile app mentioned previously in RQ1 and RQ2 that helped to achieve the learning outcomes planned in the courses.

## 4. Implications for practice in Engineering Education

In this study, four overall key elements for engineering education were entailed. The first one is the usage of in-home laboratories that according to the results evidenced can promote learning by drawing on other learning spaces different from the traditional classrooms in HEIs. This fact could enhance and reinforce the learning of the students because experimentation and hands-on tasks could be performed directly at home. Moreover, many of the course topics that may be left out due to time constraints can be reinforced with these kinds of labs. The incorporation of open-source hardware (OSHW) and open-source software (OSS) can benefit and auspicate the learning and understanding of the topics, as well as reduce the cost of the materials sent to the home of the students. A second element that deserves to be mentioned is the integration of student-created videos to evidence learning and application of the concepts by the students. While typically in prior studies the focus is on the contents in form of quizzes, lectures, and exams, with the in-home laboratories and the student-created videos the students can create other forms of expression that evidence learning but also allow them to reflect on their learning process. In addition, student-created videos and blogs allow the students to learn about ICT handling and content curation to select, adequate and study relevant information in function of the topics in this case with electronics, PhyC, basic robotics, and programming. This form of video could influence the increase of self-efficacy, but it depends on the monitoring and feedback of the teacher. A third element was the incorporation of a mobile learning app to perform PhyC and programming activities. With this application, engineering topics were addressed not only as theoretical ones but with experimentation. Then, the app and its features allowed the students to learn to program and bring students who did not like programming closer to learning the course topics which increased their motivation and self-efficacy and reduced anxiety. Even, the incorporation of PhyC activities can reduce the gender gap regarding confidence in programming that has been exposed by other prior studies. Although the primary end of this study was not to close a gender gap, the incorporation of PhyC activities with Arduino or the App can be beneficial for the academic performance of the men and women students in engineering. Finally, as a fourth element, the continuous monitoring and feedback of the teacher can benefit the learning of the students. Nonetheless, monitoring and feedback are only one of the aspects because the methodology was planned in function of the ICDM. The incorporation of activities that allow the attainment of the learning outcomes planned in the courses and the formative assessment complemented the previous element. Considering these elements is possible to create AL methodologies that can influence learning, motivation, self-efficacy, and other constructs that want to be explored in engineering education with online, blended, or mobile learning.

## 5. Limitations of the study

The first limitation is regarding the sample size of students ($n$=56) and the selection of two engineering programs that do not allow to do generalizations of the results of the methodology. However, the analysis of the results through the selected instruments shows the impacts and implications of the methodology in engineering courses with different educational characteristics which can be useful for researchers, and practitioners interested in blended or online learning methodologies in engineering. The adopted mixed approach and the triangulation of different sources to collect data provided robustness to the study. A second limitation related to the previous is that a quasi-experiment was not possible due to the number of students in each course. Nonetheless, the results were compared between courses using online and mobile learning. In addition, to gain robustness as indicated several sources of information were triangulated such as surveys, academic grades, and semi-structured interviews. A third limitation that is not discarded is the influence of a particular teaching method which could lead to the results mentioned in the RQs. The collected data evidence a predominance of the teacher characteristics in the learning process as is expected in an AL methodology. Indeed, the students evaluated anonymously the performance of the teacher in the three courses in factors such as ICT usage, pertinence in the contents, facilitation of the learning process, and openness, among others, with a mean and standard deviation of ($M = 3.84$, $SD = .214$) in a range of 0 to 4 by the end of each semester. Then, the performance of the teacher was evaluated as excellent even after the students were evaluated in the courses. This can support that the integration of the teacher characteristics, the in-home laboratories, and the student-created videos with the blogs led to a true impact on learning, motivation, and self-efficacy in the students. A four limitation is regarding the thematic analysis with the codes indicated. A single researcher made the coding of the themes in the study and under this consideration, an Intercoder Agreement (ICA) coefficient was not possible. To deal with this limitation, all information was organized, coded, and analyzed in *NVIVO* with a rigorous content-driven approach that generated the codes indicated. Finally, concerning the *EmDroid* app, it was not the purpose of this work to study how a particular set of characteristics of this influenced the PhyC or programming skills. The app was provided to the students with holistic features such as hardware interaction or block-based programming. Nevertheless, the analysis of the results indicated some special features of the app that can influence, for instance, learning, motivation, or experimentation. Further studies can be needed to analyze in-depth if a particular feature of the app can influence the learning of programming or in other constructs, e.g., self-efficacy, perceived expectancy, etc.

**6. Conclusions**

This study presented insights into an active learning methodology in online environments for engineering that was conceived considering several guidelines of the ICDM to create significant learning experiences. The active learning methodology included three overall elements: in-home laboratories, student-created videos, and teacher support and feedback. These elements interact with each other to provide the students with a way to learn and experiment with the concepts and topics related to electronics, PhyC, programming, and basic robotics. The methodology has arisen according to the lack of empirical studies on active learning and experimentation for engineering education in online educational settings, which mainly are focused either on social sciences or computer science. In this sense, with the methodology reported, it is feasible that the students to perform hands-on activities in online environments combining mobile learning, which according to the results fostered their engagement and learning process in programming and PhyC. This could be a novel initiative since it seems that online environments are more focused on a content approach instead a practical one that promotes active learning and hands-on activities.

The analysis of the collected data suggests that the incorporation of hands-on activities through the in-home laboratories in electronics and PhyC is a catalyst for the learning of the students, and it allows to foster motivation, self-efficacy, and the reduction of anxiety in the courses. These elements were pursued due to the context of the COVID-19 pandemic and to offer a real methodology for the students in online settings and not only the fast adaptation of content delivery in a format of emergency remote teaching (ERT). Similarly, the incorporation of student-created videos and blogs gave the students other forms of expression and evaluation in the courses, and these types of digital media were a source to evidence skills performance in the courses, as well as provided reflection to the students about their learning process which can be associated to certain traits of metacognitive skills according to the Marzano's taxonomy. Besides, the students reinforced their skills in editing, scripting, content curation for the information and videos in the blogs, and ICT skills.

The incorporation of mobile learning with the inclusion of the *EmDroid* app is adequate for most of the students and influenced their academic performance and their perception, e.g., on learning because the app had certain features such as affordability, integration of block-based programming, and experimentation with hardware devices. Regarding block-based programming, this allowed to bring closer students with difficulties in programming to the topics in the courses and was a way to engage students in the concepts related to PhyC and programming. Additionally, the app provides ubiquitous learning because the students can take advantage of other spaces of learning, for instance, in the home. Finally, the role of the teacher as facilitator and scaffolder of the learning process was a reiterative theme in the comments of the students through thematic analysis. Certain characteristics of the teacher were explored in this last analysis such as openness, willingness, feedback, or recursion. Also, data suggest that the role of the teacher is more accepted by women students and helps in part to reduce the difference in the academic performance between women and men students in the courses that include PhyC and programming. These results can help researchers to understand not only the technology factors in online environments for engineering education but the interaction and communication between students and teachers as a crucial factor inside this type of learning environment. In this way, the previous elements indicate that is possible to bring active learning into online educational settings for engineering.

Further work will be focused on extending the implications and refinements of the methodology in other courses of engineering as well as analyze the key features of the mobile learning app *EmDroid* for learning electronics, PhyC, and programming.

**Supplementary Materials**

Supplementary materials for the study can be found at the Zenodo Repository:

https://doi.org/10.5281/zenodo.8193140